\documentclass[12pt]{article}
\usepackage{amssymb}

\usepackage{amsmath}

\begin{document}

\begin{center}
\vspace{1cm}{\Large {\bf Free Field Equations For Space-Time
Algebras With Tensorial Momentum}}

\vspace{1cm} {\bf R. Manvelyan} \footnote{ E-mail:
manvel@moon.yerphi.am}  and {\bf R. Mkrtchyan} \footnote{ E-mail:
mrl@r.am} \vspace{1cm}

\vspace{1cm}

{\it Theoretical Physics Department,} {\it Yerevan Physics
Institute,}

{\it Alikhanian Br. St.2, Yerevan, 375036 Armenia }
\end{center}

\vspace{1cm}
\begin{abstract}
Free field equations, with various spins, for space-time algebras
with second-rank tensor (instead of usual vector) momentum are
constructed. Similar algebras are appearing in superstring/M
theories. The most attention is payed to the gauge invariance
properties, particularly the spin two equations with gauge
invariance are constructed for dimensions 2+2 and 2+4 and
connection to Einstein equation and diffeomorphism invariance is
established.
\end{abstract}

\renewcommand{\thefootnote}{\arabic{footnote}} \setcounter{footnote}0
{\smallskip \pagebreak }

\section{Introduction\newline}

        We will consider unitary field theories
with space-time symmetry algebra, differing from usual Poincare
algebra and consisting from two sets of generators  $M_{\mu \nu }$
and
$Z_{\mu \nu }$  combined into semidirect product of $so(p,q)$ ($%
M_{\mu \nu }$) and an Abelian group of second-rank antisymmetric
tensors $Z_{\mu \nu }$. Thus the algebra is:
\begin{equation}
\begin{array}{l}
[M_{\mu \nu } ,M_{\lambda \sigma } ] = \eta _{\mu \lambda } M_{\nu
\sigma }  - \eta _{\nu \lambda } M_{\mu \sigma }  + \eta _{\nu
\sigma } M_{\mu \lambda }  - \eta _{\mu \sigma } M_{\nu \lambda } \label{mmz} \\
\left[M_{\mu \nu } ,Z_{\lambda \sigma } \right] = \eta _{\mu
\lambda } Z_{\nu \sigma }  - \eta _{\nu \lambda } Z_{\mu \sigma }
+ \eta _{\nu
\sigma } Z_{\mu \lambda }  - \eta _{\mu \sigma } Z_{\nu \lambda }\\
\left[Z_{\mu \nu } ,Z_{\lambda \sigma }\right]=0
\end{array}
\end{equation}
where the signature is $\eta_{\mu\nu}=(+,..+,-,-,-...)$ and
indices are running over $d=p+q$ time+space values. This algebra
differs from Poincare one in two respects: first, role of
translation generators are now played by second-rank tensor, not
vector, and, second, signature of metric is different. Actually we
shall concentrate on a specific values of p, q, (p=2,
particularly) but some considerations will be applicable to more
general cases.
    This algebra is inspired by development of superstring/$M$ theory. In the
supersymmetry algebras of that theories \cite{F} p=1, q=10,9,...
and tensorial charges (branes charges) appear, particularly
$Z_{\mu \nu}$, as well as other tensors. The most general is the
M-theory supersymmetry algebra at d=11, with the anticommutator of
supercharges equal to
\begin{eqnarray}
\left\{ \bar{Q},Q\right\} &=&\Gamma ^{i}P_{i}+\Gamma
^{ij}Z_{ij}+\Gamma
^{ijklm}Z_{ijklm},  \label{1} \\
i,j,... &=&0,1,2,..10  \notag
\end{eqnarray}
so our algebra is particular case when  only $Z_{\mu \nu}\ \ne\
0$, and $p=1, q=10$. Energy-momentum $P_{\mu}$ naturally
disappeared the other case, with p=2, q=10. This algebra appeared
first time in \cite{Bars} as a possible generalization of
space-time symmetry algebra of M theory. Namely, interpreting
Majorana spinor Q as a Majorana-Weyl with respect to 2+10
dimensional Lorentz group, combining $P_{i}$ and $Z_{ij}$ into one
12d tensor $P_{\mu \nu }$, and interpreting $Z_{ijklm}$ as a
self-dual 12d sixth-rank antisymmetric tensor, we obtain:

\begin{eqnarray}
\left\{ \bar{Q},Q\right\} &=& \Gamma ^{\mu \nu }P_{\mu \nu
}+\Gamma
 ^{\mu \nu \lambda \rho \sigma \delta }Z_{\mu \nu
\lambda \rho \sigma \delta }^{+}
\label{2} \\
\mu \nu ,... &=&0^{\prime },0,1,...10  \notag
\end{eqnarray}
with (\ref{mmz}) as a bosonic subalgebra of corresponding
superalgebra. Let's stress that it is not possible to add a vector
$P_{\mu}$ in the r.h.s., due to the properties of gamma-matrixes.
So, the study of algebra (\ref{mmz}) can be useful, in different
ways of study of superstring/M theory.
    The dimensions higher than 11 and different
supersymmetry algebras appeared many times in the study of
M-theory. Papers \cite{list} represent some of that
investigations.

In this paper we shall restrict ourselves mainly to purely bosonic
case (\ref{mmz}), without fermionic part (\ref{2}).

Our aim is, taking seriously algebra (\ref{mmz}), to construct an
interacting field theories with such a symmetry algebras. Like for
standard Poincare algebra, the first step to interacting field
theories may be the construction of relativistic field equations.
This is another way of the description of  the unitary
representations of the algebra, which is adapted for the further
construction of the invariant interacting theories, since all
elements in that field equations are covariant. As is well-known,
one of the main features for such equations is the gauge
invariance, which is responsible for removal of superfluous fields
components. In previous paper \cite{Man1} we constructed field
equations for scalar, spinor and vector fields. Particularly, it
was found the adequate definition of gauge invariance. In this
paper we shall consider mainly the free spin two field.
    In Sect. 2 we recall some facts from the \cite{Man1}
   on the
field equations and gauge invariance for the algebra (\ref{mmz}).
In Sect.3 the gauge-invariant gravitino equation in dimension 2+2
is presented. Sections 4, 5  and 6 are devoted to the free gravity
equations in 2+q dimensions, and their gauge invariance properties
in dimensions 2+4 and 2+6. Conclusion contains discussion of the
possible development of present theory.

\section{Basic elements of d=2+q theory}
    The important feature of (\ref{mmz}) is the presence of many
invariants (Casimir's operators), constructed from $P_{\mu\nu}$:
\begin{eqnarray}
TrP^{2} &=&P_{\mu \nu }P^{\nu \mu },  \notag \\
TrP^{4} &=&P_{\mu \nu }P^{\nu \lambda }P_{\lambda \rho }P^{\rho \mu },
\label{3} \\
&...&  \notag
\end{eqnarray}
instead of the single one in algebras with momentum only:
$P^{2}=P_{i}P^{i}$. This means, that now, even in the simplest
case of scalar theory,we have to introduce  few equations of
motion instead of the one Klein-Gordon:

\begin{eqnarray}
(TrP^{2}-2m_{1}^{2})\Phi (P_{\mu \nu }) &=&0,  \label{4} \\
(TrP^{4}-2m_{2}^{4})\Phi (P_{\mu \nu }) &=&0,  \nonumber \\
&&...  \nonumber
\end{eqnarray}

    We will call these equations respectively "first level",
"second level", etc.
    Evidently, in these equations $\Phi$ is the scalar function,
Fourier transform of which depends on a coordinates $x^{\mu \nu}$
conjugate to momenta $P_{\mu \nu}$. So, these equations, as well
as others below, are differential equations in the $d(d-1)/2$
dimensional "space-time" with coordinates $x^{\mu\nu}$.

According to the Wigner's  little group's method of construction
of unitary representations of such a semidirect products, we have
to choose an orbit (\ref{4}) of a Lorentz group in the space
$P_{\mu \nu}$, take some arbitrary point on that orbit and then
find a stabilizer of that point (=little group). The unitary
representations of little group are giving rise to the unitary
representations of initial group, through the procedure of
induction. The special orbit is one with all $m_{i}=0$. For p=2,
q=10 this orbit corresponds to the massless multiplet of d=11
supergravity. The possible choice of point on this orbit is:
\begin{eqnarray}
P_{\mu \nu } &=&(P_{0^{\prime }i},P_{ij}=0),  \label{41} \\
P^{2} &=&P_{0^{\prime }i}P^{0^{\prime }i}=0,  \label{5} \\
P^{0^{\prime }i} &=&(1,1,0,...,0)  \label{6}
\end{eqnarray}
    The Lie algebra of little group of this orbit can be found as
follows. Consider all elements of $so(2,q)$ algebra, which leave
the tensor $(\ref{4})$ unchanged. Actually (\ref{4}) itself, with
raised second index is an element of $so(2,q)$, so we are seeking
its stabilizer in this algebra. It is easy to show, that the
following matrices of $so(2,q)$ are exactly all matrices,
commuting with $(\ref{4})$:
\begin{equation}
\left(
\begin{array}{ccccccc}
0 & a & a & 0 & 0 & ... & 0 \\
-a & 0 & 0 & b & c & ... & d \\
a & 0 & 0 & -b & -c & ... & -d \\
0 & b & b & 0 & e & ... & f \\
0 & c & c & -e & 0 & ... & g \\
... & ... & ... & ... & ... & ... & ... \\
0 & d & d & -f & -g & ... & 0
\end{array}
\right)  \label{m1}
\end{equation}

The algebra of matrixes $(\ref{m1})$ is a direct sum of the
$so(1,1)$ algebra of matrices $(\ref{m1})$ with only non-zero
entry $a$, and an algebra, which is a semidirect sum of $so(d-3)$
(represented by matrixes (\ref {m1}) with $a=b=c=...=d=0$) and an
Abelian algebra of matrices $(\ref{m1})$ with non-zero elements
$b,c,...,d$ only. The unitary finite-dimensional representations
of this little group algebra are those of $so(d-3)$ subalgebra,
with other generators represented by zero, which is possible due
to the structure of the algebra.

The role of equations (\ref{4}) is to bring the general function
of the momenta $ P_{\mu \nu }$ to a function on the orbit. For
non-trivial representations of little group the relativistic
fields contain additional components, which can be removed by
gauge transformations. The notion of gauge symmetry generalizes in
this case, as shown in \cite{Man1}, to the single equation that a
sum of variations of these actions is equal to zero.

The first example of fields with gauge invariance is the vector
field theory. In momentum representation, introducing the vector
field $A_{\mu }(P_{\lambda \nu })$ and the ``field strength''
\begin{equation}
F_{\mu \nu \lambda }=P_{\mu \nu }A_{\lambda }+P_{\nu \lambda
}A_{\mu }+P_{\lambda \mu }A_{\nu }  \label{14}
\end{equation}
we can define the following set of equations of motion:
\begin{eqnarray}
P^{\mu \nu }F_{\mu \nu \lambda } &=&0  \label{15-0.1} \\
(P^{3})^{\mu \nu }F_{\mu \nu \lambda } &=&0  \label{15-0.2} \\
&&... \\
(P^{d-1})^{\mu \nu }F_{\mu \nu \lambda } &=&0  \label{15-0.4}
\end{eqnarray}
An equivalent set of equations is:
\begin{eqnarray}
P^{\mu \nu }F_{\mu \nu \lambda } &=&2\left( P_{\lambda }^{2\mu
}-\frac{1}{2}
TrP^{2}\delta _{\lambda }^{\mu }\right) A_{\mu }=0  \label{15.1} \\
G_{2i}(P)A_{\mu } &=&0\,\,\,\,\,\,\,i=1,2,3,...  \label{15.2}
\end{eqnarray}
Here $G_{i}$  are combinations of invariants (\ref{4}) coinciding
with the coefficients in the expansion of the characteristic
polynomial for matrix $P_{\mu }^{\quad \nu }$:
\begin{eqnarray}
f(x) &=&\det (P-x)  \label{15.3} \\
&=&x^{d}+x^{d-2}G_{2}+x^{d-4}G_{4}+...  \\
G_{2}&=&-\frac{1}{2}TrP^{2}
\end{eqnarray}
The form of  $G_{i}$ is independent of dimensionality $d$. Again
this set is equivalent to the ordinary Maxwell system under the
condition $P_{ij}=0$ and therefore reproduces the masslessness
condition $ P^{2}=0$.

We can define the set of actions corresponding to each set of
equations. For  $(\ref{15.1})$, $(\ref{15.2})$ these are:
\begin{eqnarray}
S_{1} &=&\int [dX^{\mu \nu }]\left( -\frac{1}{6}F^{\mu \nu \lambda
}F_{\mu
\nu \lambda }\right) ,  \label{15.4} \\
S_{i} &=&k_{2i}\int [dX^{\mu \nu }]\left( \frac{1}{2}A_{\lambda
}G_{2i}(\partial _{X^{\mu \nu }})A^{\lambda }\right) ,  \label{15.5} \\
i &=&1,2,..  \nonumber
\end{eqnarray}
Here we introduced coupling constants $k_{2i}$ with appropriate
dimensionality. For study of gauge invariance properties of this
theory, let's define the following variation of field A$ _{\mu }$
(take d=2+10, for definiteness):
\begin{equation}
\delta A_{\mu }(P_{\lambda \rho })=P_{\mu \nu }\alpha ^{\nu
}\delta (G_{2}(P))...\delta (G_{6}(P))  \label{d1}
\end{equation}

Eqs. of motion (\ref{15.2}) are evidently invariant with respect
to this transformation, the (\ref{15.1}) is also invariant,
because delta-functions in ( \ref{d1}) put $P_{\mu \nu }$ into the
form (\ref{4}), and (\ref{15.1}) produces usual Maxwell equation
with usual gauge transformation (\ref{d1}). It is easy to see,
that gauge transformation (\ref{d1}) permits one to gauge away $
A_{0^{^{\prime }}}$ component and longitudinal part of $A_{i}$
(taking into account that $A_{\mu }$ is non-zero only on the shell
of delta-functions $ \delta (G_{2}(P))$, ...,$\delta (G_{6}(P))$):
\begin{eqnarray}
\delta A_{0^{\prime }}(P_{0^{\prime }i}) &=&P_{0^{\prime }\nu
}\alpha ^{\nu
}(P_{0^{\prime }i})=\beta  \label{d2} \\
\delta A_{i}(P_{0^{\prime }i}) &=&P_{i\nu }\alpha ^{\nu
}(P_{0^{\prime }i})=P_{i0^{\prime }}\gamma  \label{d3}
\end{eqnarray}
So, these equations describe the vector representation of SO(9)
group, as desired. Note that actions (\ref{15.4}), (\ref{15.5})
are invariant, also.

Another way of realization of the same idea of gauge invariance is
the following. We can define the following set of variations
corresponding to the each equation of the set:
\begin{eqnarray}
\delta _{1}A_{\mu }(P_{\lambda \rho }) &=&P_{\mu \nu }^{10}\alpha
^{\nu
}(P_{\lambda \rho })  \nonumber \\
\delta _{2}A_{\mu }(P_{\lambda \rho }) &=&k_{2}^{-1}P_{\mu \nu
}^{8}\alpha
^{\nu }(P_{\lambda \rho })  \label{16} \\
&.&  \nonumber \\
&.&  \nonumber \\
\delta _{6}A_{\mu }(P_{\lambda \rho }) &=&k_{6}^{-1}\alpha _{\mu
}(P_{\lambda \rho })  \nonumber
\end{eqnarray}
Then it is easy to see that the following equation is satisfied:
\begin{equation}
\delta _{1}S_{1}+\delta _{2}S_{2}+....+\delta _{6}S_{6}=0
\label{17}
\end{equation}
due to the well-known Hamilton-Cayley identity for characteristic
polynomials:
\begin{equation}
f(P)=0  \label{18}
\end{equation}
where
\begin{equation}
f(x)=\det (P-x)  \nonumber
\end{equation}
As above, assuming that higher eqns. $\delta S_{i}/A_{\mu }$
$(i=2,...,6)$ are satisfied, eqn. (\ref{17}) gives the usual
statement of gauge invariance of Maxwell's action. (The equation
similar to (\ref{17}) is valid also for other sets of actions,
with different variations $\delta _{i}A_{\mu }$.) Then, under the
condition $P_{ij}=0$ the remaining symmetry is:
\begin{eqnarray}
\delta A_{0^{\prime }}(P_{0^{\prime }i}) &=&P_{0^{\prime }\nu
}^{10}\alpha
^{\nu }(P_{0^{\prime }i})=\beta  \label{19} \\
\delta A_{i}(P_{0^{\prime }i}) &=&P_{i\nu }^{10}\alpha ^{\nu
}(P_{0^{\prime }i})=P_{i0^{\prime }}\gamma  \label{20}
\end{eqnarray}
The first one can be used for gauging away the additional twelfth
component $ A_{0^{\prime }}$, the second one gives the usual gauge
transformation for the remaining $11d$ Abelian gauge field. The
subtlety in (\ref{19}), (\ref{20}) is that: if we consider the
on-shell condition $P^{2}=0$ from the beginning, it is not
possible to gauge away $A_{0^{\prime }},$ as is seen from
(\ref{19}).

Let's finish this section with few remarks. First one is that
there is another possibility for gauge-invariant actions. The last
equation $\det (p) = 0$ in (\ref{15.2}) can be replaced by
equation $Pfaff(p)h = 0$, where
$Pfaff(p)=\epsilon_{\mu\nu...\alpha\beta}p^{\mu\nu}...p^{\alpha\beta}$.
It is possible to find corresponding variations of fields in that
set of equations such that Eq.(\ref{17}) again will be satisfied.

Next,  Eqn. (\ref{17}) in this form is appropriate for
generalization to non-quadratic Lagrangians, but we are not aware
on any general considerations proving that such equations are
direct consequence of a multilagrangian nature of the theory.
Moreover, taking into account that the main sense of the equation
(\ref {17}) is that when fields are satisfying the second and
higher level equations of motion, then (\ref {17}) reduces to
standard equation of symmetry of first action, one can suggest
other, nonlinear relations between variations of actions, such
that variation of first one is zero when others are zero. E.g.
$\delta _1 S_1  = \sqrt{(\delta _2 S_2 )} + ...$. In this
expression parameters of infinitesimal variations are implied to
be removed from variations.

\bigskip

\section{Free gravitino field.}
    Next field with gauge invariance can be the gravitino field.
    Equations are given in \cite{Man1}.
We shall consider that in dimension  $d=(2+2)$:
\begin{eqnarray}
\gamma ^{\mu \nu \lambda \rho }\partial _{\nu \lambda }\psi _{\rho}=0\\
 Pfaff(p)\psi _{\mu }=0
\end{eqnarray}
Gauge transformations parameter, as in the vector case, has one
more vector index in comparison with case of usual Rarita-Shwinger
field, i.e. is spin-vector:
\begin{eqnarray}
\delta _{1}\psi _{\rho }=\partial _{\rho \nu }\varepsilon _{\nu
}\\
\delta _{2}\psi _{\nu }\backsim \varepsilon _{\nu }
\end{eqnarray}
Check of gauge invariance (\ref{17}):
\begin{equation}
\delta _{1}(\gamma ^{\mu \nu \lambda \rho }\partial _{\nu \lambda
}\psi _{\rho })=\gamma ^{5}\widetilde{p}^{\mu \rho }p_{\rho
\upsilon }\varepsilon _{\nu }\backsim Pfaff(p)\varepsilon _{\mu
}=\delta _{2}(Pfaff(p)\psi _{\mu })
\end{equation}

After reduction, gauge transformations give the gauge invariances
of reduced equation ($(1+2)$ gravitino equation)

\begin{equation}
\delta _{1}\psi _{i}=-p_{i}\varepsilon _{0^{\prime }}
\end{equation}

which coincide with gravitino gauge transformations. Remaining
part of gauge invariance
\begin{equation}
\delta _{1}\psi _{0^{\prime }}=p_{k}\varepsilon _{k}
\end{equation}

permits one to gauge away $\psi _{0^{\prime }}.$

\section{Free Gravity}

    We suggested in \cite{Man1} the following quadratic first level Lagrangian
for the symmetric tensor field $h_{\mu \nu }$:

\begin{equation}
\begin{array}{l}
L_{1}=-\frac{1}{4}h_{\mu \nu }(\partial _{\lambda \rho }\partial
^{\rho \lambda })h^{\mu \nu } -\frac{1}{2}h_{\mu \nu }(\partial
^{\mu \lambda }\partial ^{\nu \rho })h_{\lambda \rho }\\
+h_{\mu
\nu }(\partial ^{\mu \lambda }\partial _{\lambda \rho })h^{\rho
\nu }-h_{\mu }^{\nu }(\partial ^{\mu \lambda }\partial _{\lambda
\nu })h+ \frac{1}{4}h(\partial ^{\mu \nu }\partial _{\nu \mu })h
\label{L2}
\end{array}
\end{equation}
This expression is unique among those of second order over
derivatives and over field $h_{\mu \nu }$, which goes into
quadratic part of General Relativity Lagrangian after reduction
(\ref{41}). That contains no additional degrees of freedom. More
exactly, reduction means that fields are independent of
coordinates $x^{ij},$ and we shall use notations:

\begin{eqnarray}
\mu =(0^{^{\prime }},i) \\
h_{0^{^{\prime }}0^{^{\prime }}}=a,\\
h_{0^{^{\prime }}i}=b_{i},\\
 h_{\mu }^{\mu} =h, \\
\partial
_{i}=\partial _{0^{^{\prime }}i} \\
h_{i}^{i}=u, \\
h=a+u
\end{eqnarray}
After reduction $L_{2}$ goes into

\begin{equation}
\widetilde{L}_{1}=-\partial _{i}h_{ij}\partial _{j}u+(1/2)\partial
_{i}u\partial _{i}u+\partial _{i}h_{ij}\partial
_{k}h_{kj}-(1/2)\partial _{k}h_{ij}\partial _{k}h_{ij}  \label{l2}
\end{equation}

which is quadratic part of Einstein-Gilbert Lagrangian. Note that
components $ h_{0^{^{\prime }}0^{^{\prime }}}=a,h_{0^{^{\prime
}}i}=b_{i}$ disappeared.

Corresponding to (\ref{l2}) equation of motion is (in momentum
representation):

\begin{eqnarray}
E_{\mu \nu }-\frac{1}{3}E_{\lambda }^{\lambda }\eta _{\mu \nu }=0
\label{Eq} \\
E_{\mu \nu } =(-1/2)p_{\alpha \beta }p^{\beta \alpha }h_{\mu \nu
}+(php)_{\mu \nu }+(p^{2}h+hp^{2})_{\mu \nu }-(p^{2})_{\mu
\nu}h_{\lambda}^{\lambda }
\end{eqnarray}

So, at $d=3$ equation is trivial, which is in agreement with
desired correspondence of this theory with trivial $d=2$ Einstein
gravity. Reduction of Eq.(\ref{Eq}) gives the linearized Einstein
equation

\begin{equation}
p^{2}h_{ij}+p_{i}p_{j}h_{k}^{k}-p_{i}p_{k}h_{j}^{k}-p_{j}p_{k}h_{i}^{k}=0
\label{eq}
\end{equation}

and nothing else, in agreement with reduction of Lagrangian
(\ref{L2}). It is reasonable to assume, that (\ref{L2}) has a
gauge invariance, which after reduction removes $a$ and $b_{i}$
and becomes a usual diffeomorphism invariance of (\ref{l2}) and
(\ref{eq}). That is discussed in next Sections.

\section{Gauge Invariance of Free Gravity at d=2+2 }

For this dimensionality we have to add only one, second level,
Lagrangian to (\ref{L2}):

\begin{eqnarray}
S_2=\int L_{2} =\frac{1}{2}\int h_{\mu \nu }Pfaff(\partial)h_{\mu \nu } \label{22eq}\\
\frac {\delta S_2} {\delta h_{\mu \nu }}=K^{\mu \nu }
=Pfaff(p)h^{\mu \nu }=0
\end{eqnarray}

where
\begin{eqnarray}
Pfaff(p) &=&p_{\mu \nu }p_{\lambda \sigma }\varepsilon ^{\mu \nu
\lambda
\sigma }=\widetilde{p}_{\mu \nu }p^{\mu \nu } \\
\widetilde{p}_{\mu \nu } &=&\varepsilon _{\mu \nu \lambda \sigma }p^{\lambda
\sigma } \\
\det (p) &=&(1/64)(Pfaff(p))^{2}
\end{eqnarray}

This theory has the following two gauge invariances
\begin{eqnarray}
\delta _1 S_1  + \delta _2 S_2  = 0 \label{22}
\end{eqnarray}
with
\begin{eqnarray}
\delta _{1}h_{\mu \nu } &=&(p^{2}\xi +\xi p^{2})_{\mu \nu } \label{221}\\
\delta _{2}h_{\mu \nu } &=&\frac{1}{4}( \widetilde{p}\xi p+p\xi
\widetilde{p})_{\mu \nu }
\end{eqnarray}
and
\begin{eqnarray}
\delta _{1}h_{\mu \nu } &=&(p\sigma p)_{\mu \nu } \label{222}\\
\delta _{2}h_{\mu \nu } &=&-\frac{1}{64}\sigma _{\mu \nu }Pfaff(p)
\end{eqnarray}
where $\xi_{\mu \nu }$ and $\sigma_{\mu \nu }$ are symmetric
tensors, parameters of transformation. The reduction of these
gauge invariances gives the standard gauge invariance of
$\widetilde{L}_{1}$ and possibility to gauge away superfluous
components of $h_{\mu \nu }$:
\begin{eqnarray}
\delta h_{0^{\prime }0^{\prime }}&=&-2p^{2}\xi _{0^{\prime
}0^{\prime }}-p^{i}p^{j}\sigma_{ij}\\
 \delta h_{0^{\prime
}k}&=&-p^{2}\xi _{0^{\prime }k}-p_{k}( p^{l}\xi _{0^{\prime
}l})+p_{k}( p^{l}\sigma _{0^{\prime}l})\\
 \delta h_{ij}&=&-p_{i}p^{k}\xi
_{kj}-p_{j}p^{k}\xi _{ki}-p_{i}p_{j}\sigma _{0^{\prime }0^{\prime
}}
\end{eqnarray}

\bigskip

\section{d=2+4 Free Gravity\newline}

    Lagrangian and equation of motion for field $h_{\mu \nu }$, which
reduces to Einstein ones are given above - (\ref{L2})(\ref{Eq}).
Next level Lagrangians and equations, in analogy with above, can
be assumed as:
\bigskip
\begin{eqnarray}
G_{\mu \nu } &=&G_{4}h_{\mu \nu }=0  \label{Eq62} \\
K_{\mu \nu } &=&Pfaff(p)h_{\mu \nu }=0  \label{Eq63}
\end{eqnarray}
The problem is in gauge invariance. It should be of the following
form:

\begin{eqnarray}
\delta _1 S_1  + \delta _2 S_2  + \delta _3 S_3  = 0 \\
 \delta_{1}E_{\mu \nu }+\delta _{2}G_{\mu \nu }+\delta _{3}K_{\mu \nu }=0
\end{eqnarray}
with some variations $\delta _{i},$ $i=1,2,3$ of field
$h_{\mu\nu}$.

We have an analog of (\ref{222}):
\begin{eqnarray}
\delta _{1}h &=&p^{2}\xi p^{2}-\frac{1}{3}p^{2}tr(p^{2}\xi ) \label{var62}\\
\delta _{2}h &=&-p\xi p -\frac {1}{3}tr(p^2\xi)\\
\delta _{3}h &=&\frac{1}{384}(p\xi q+q\xi
p)-\frac{1}{384}Pfaff(p)(\xi)
\end{eqnarray}
where

\begin{equation}
q_{\mu \nu }=\epsilon _{\mu \nu \lambda \sigma \rho \tau }p^{\lambda \sigma
}p^{\rho \tau }
\end{equation}

After reduction this invariance gives the following invariance of
Einstein equation:

\begin{equation}
\delta h_{ij}\backsim p_{i}p_{j}\rho
\end{equation}
where $\rho $ is constructed from $p$ and $\xi$. This is only a
part of diffeomorphism invariance of Einstein equation.

Simultaneously, we argue, that there exist gauge invariance of the
first kind (analog of (\ref{221})), which is already sufficient
for all purposes on removing the superfluous components, and which
transforms into the full diffeomorphism invariance of Einstein
equation. This can be seen from the following. Let's assume the
"diagonalized" form of matrix $p$:

$p=
\begin{array}{cccccc}
0 & a & 0 & 0 & 0 & 0 \\
-a & 0 & 0 & 0 & 0 & 0 \\
0 & 0 & 0 & b & 0 & 0 \\
0 & 0 & -b & 0 & 0 & 0 \\
0 & 0 & 0 & 0 & 0 & c \\
0 & 0 & 0 & 0 & -c & 0
\end{array}
\allowbreak $

The assumed gauge transformation is

\begin{equation}
\delta h=p^{4}\xi +\xi p^{4}  \label{var61}
\end{equation}
where $\xi_{\mu\nu}$ is symmetric second rank tensor. Variation of
equation of motion (\ref{Eq}) gives (we present only first column
of equation, due to lack of space):
\begin{equation}
\begin{array}{*{20}c}
   2b^{2}a^{4}\xi_{11}+2c^{2}a^{4}\xi_{11}+2a^{2}b^{4}\xi_{33}+2a^{2}b^{4}\xi_{44}+2c^{4}\xi_{55}a^{2}+2a^{2}c^{4}\xi_{66}  \\
    2a^{4}\xi_{12}b^{2}+2a^{4}\xi_{12}c^{2} \\
  -a^{5}b\xi_{24}-ab^{5}\xi_{24}+c^{2}a^{4}\xi_{13}+c^{2}b^{4}\xi_{13}  \\
   a^{5}b\xi_{23}+ab^{5}\xi_{23}+c^{2}a^{4}\xi_{14}+c^{2}b^{4}\xi_{14}  \\
   -a^{5}c\xi_{26}-ac^{5}\xi_{26}+b^{2}a^{4}\xi_{15}+b^{2}c^{4}\xi_{15} \\
   a^{5}c\xi_{25}+ac^{5}\xi_{25}+b^{2}a^{4}\xi_{16}+b^{2}c^{4}\xi_{16}  \\
\end{array}
\end{equation}

As is easily seen, this expression is equal to zero when two from
three ''eigenvalues'' $a, b$ or $c$ are equal to zero. This last
condition follows from  equations  (\ref{Eq62}), (\ref{Eq63}). So,
the variation of Lagrangian (\ref{L2}) is zero when variations of
two others are zero. Unfortunately, we didn't find such a
relation. The problem can be formulated as follows: 6d relation
has to pass into 4d relation when all objects are restricted to
4d, tentative 6d relation has a $ G_{4}$ which goes into
$det_{4}(p)$ when restricted to surface $det_{6}(p)=0$, but in 4d
gauge invariance relation (\ref{22}) enters Pfaffian, not
determinant.

The solution we are suggesting for this problem  is the following:
we change the second level action $S_2$, writing instead
\begin{equation}
\begin{array}{l}
 {\rm S}_{\rm 2}  = \int {(dx)(c(h_{\mu \nu } \partial ^4 _{\nu \lambda } h_{\lambda \mu } }  - \frac{1}{2}h_{\mu \lambda } \partial ^2 \partial ^2 _{\mu \nu } h_{\nu \lambda } )) +  \label{24}\\
 h_{\mu \sigma } \partial _{\mu \nu } \partial ^3 _{\lambda \sigma } h_{\nu \lambda }  - (h_{\mu \sigma } \partial ^2 \partial _{\mu \nu } \partial _{\lambda \sigma } h_{\nu \lambda } ) + (1 + c)hG_4 h) \\
 \end{array}
\end{equation}
And corresponding equation in momentum representation:
\begin{equation}
\begin{array}{l}
 c(p^4 h + hp^4  - \frac{1}{2}Sp(p^2 )(p^2 h + hp^2 )) +  \\
 (p^3 hp + php^3  - Sp(p^2 )php) + (1 + c)G_4 h =0\\
 \end{array}
\end{equation}
where $c = 1 \pm \sqrt 2$. This equation itself is a consequence
 of (\ref{Eq62}), (\ref{Eq63}), and vice versa: from (\ref{24}) together with
 (\ref{Eq63}) follows that two from three eigenvalues $a, b, c$
 are equal to zero. Namely, as shown below in (\ref{red}),
 (\ref{24}) goes into combination of Pfaffian and determinant
 terms under reduction.

Now we have desired gauge transformation:

\begin{eqnarray}
\delta _1 h_{\mu \nu }  &=& (\partial ^4 \xi)_{\mu \nu }  + (\mu
\leftrightarrow \nu ) - \frac{2}{3}\partial ^2 _{\mu \nu }
\partial ^2 _{\alpha \beta } \xi_{\alpha \beta }\\
\delta _2 h_{\mu \nu }  &=& (\partial ^2 \xi)_{\mu \nu }  +
(\mu  \leftrightarrow \nu ) + (1 - c)\partial _{\mu \sigma }
\partial _{\nu \lambda } \xi_{\sigma \lambda }  \\
\delta _3 h_{\mu \nu }  &=& \frac {1}{48}(1 +
c)Pfaff(\partial)(q_{\lambda \nu } \partial _{\mu \sigma }
\xi_{\sigma \lambda }  + (\mu  \leftrightarrow \nu
))\nonumber\\&+& det(\partial)(2c\xi +
 2Sp(\xi))_{\mu \nu }
\end{eqnarray}
In momentum representation:
\begin{eqnarray}
 \delta _1 h_{\mu \nu }  &=& (p^4 \xi+\xi p^4 )_{\mu \nu }  - \frac{2}{3}p^2 _{\mu \nu } Sp(p^2 \xi) \\
 \delta _2 h_{\mu \nu }  &=& (p^2 \xi + \xi p^2 )_{\mu \nu }  + (1 - c)(p\xi p)_{\mu \nu }  \\
 \delta _3 h_{\mu \nu }  &=& \frac {1}{48}(1 + c)(q\xi p + p\xi q)_{\mu \nu } + \frac {1}{48} Pfaff(p)(2c\xi + 2Sp(\xi))_{\mu \nu }
\end{eqnarray}
where
\begin{eqnarray}
q_{\mu \nu }= \frac{1}{8}\varepsilon _{\mu \nu \lambda \rho \sigma
\tau }p^{\lambda \rho}p^{\sigma \tau}
\end{eqnarray}
 The gauge transformation (\ref{var62}) doesn't disappear in
this new set of equations, instead we have invariance with respect
to:

\begin{eqnarray}
&\delta _1 h_{\mu \nu } = (p^3 hp+php^3 )_{\mu \nu}+(c-1)(p^2 hp^2 )_{\mu \nu }  - \frac{1}{3}(1 + c)p^2 _{\mu \nu } Sp(p^2 h) \\
&\delta _2 h_{\mu \nu}  = c(php)_{\mu \nu }  \\
&\delta_3 h_{\mu \nu}  = - \frac {1}{48}Pfaff(p)(1 + c)(h -
Sp(h))_{\mu \nu }
 \end{eqnarray}

These gauge transformations, as in d=2+2, give after reduction the
five-dimensional diffeomorphism transformation, as well as
possibility to exclude the superfluous components of tensor
$h_{\mu\nu}$. The other check is that when  reducing to d=2+2 we
obtain some combination of invariances (\ref{221}) (\ref{22}).
Exactly, equation (\ref{24}) under that reduction goes into
combination of Pfaffian and determinant terms (instead of pure
Pfaffian in (\ref{22eq})):
\begin{equation}
\begin{array}{l}
(p^3 hp + php^3  - Sp(p^2 )php)+(c-1)det(p)h=\\
\frac{1}{8}Pfaff(p)(phq+qhp)+(c-1)det(p)h =0 \label{red}
\end{array}
\end{equation}

Evidently, if we choose this equation in $2+2$ instead of pure
Pfaffian term, only some combination of (\ref{22}),(\ref{221})
will survive. The question is whether the interaction will prefer
this combination (if any).

\bigskip

\section{Conclusion}
Generalization of previous results to arbitrary dimension should
be possible. Another direction of development is consideration of
other orbit of algebra, particularly that corresponding to
membrane. Membrane corresponds, from the point of view of algebra
(\ref{1}), to orbit with e.g. $P_{0^{\prime }0}=m,$ $P_{12}=m$
\cite{F}. Invariants are equal to:

\begin{eqnarray}
TrP^{2} &=&-4m^{2}  \notag \\
TrP^{4} &=&4m^{4} \\
TrP^{6} &=&-4m^{6}  \notag \\
&&....  \notag
\end{eqnarray}

The algebra of little group of this orbit in dimension $d=2+(d-2)$
is $ so(2)+so(2,1)+so(d-4)$. As always, first factor corresponds
to matrix $ P_{\mu \nu }$ itself. Matrices from the little group
algebra have the form (second index raised):

\begin{equation}
\begin{array}{cccccccc}
0 & p_{12} & p_{13} & p_{14} & 0 & 0 & 0 & 0 \\
-p_{12} & 0 & -p_{14} & p_{13} & 0 & 0 & 0 & 0 \\
p_{13} & -p_{14} & 0 & p_{34} & 0 & 0 & 0 & 0 \\
p_{14} & p_{13} & -p_{34} & 0 & 0 & 0 & 0 & 0 \\
0 & 0 & 0 & 0 & 0 & -p_{56} & -p_{57} & -p_{58} \\
0 & 0 & 0 & 0 & p_{65} & 0 & -p_{67} & -p_{68} \\
0 & 0 & 0 & 0 & p_{75} & p_{76} & 0 & -p_{78} \\
0 & 0 & 0 & 0 & p_{85} & p_{86} & p_{87} & 0
\end{array}
\end{equation}

\bigskip

The right low block of matrix is algebra $so(d-4)$ (drawn for the
case $d=8$), the left upper block is the direct sum of $so(2)$
matrix (with $ p_{12}=p_{34},$other elements are zero) , and
$so(2,1)$ (with $p_{12}=-p_{34} $).

The simplest unitary representations of the little group is, as usual, the
scalar one, when group is represented trivially, then we can consider
non-trivial unitary finite representations of factors $so(2)$ and $so(d-4).$
So, for scalar field $\varphi $ equations of motion are

\begin{eqnarray}
(TrP^{2}+4m^{2})\varphi  &=&0  \notag \\
(TrP^{4}-4m^{4})\varphi  &=&0 \\
(TrP^{6}+4m^{6})\varphi  &=&0  \notag \\
&&....  \notag
\end{eqnarray}
 As in above, the other representations can be considered.

 Next important problem is construction of interaction terms. It is
 reasonable to approach this
 problem by Nether procedure, starting
 from linear equations and their gauge invariance transformations,
 then constructing next order term both in equations and in
 transformations from
 the requirement
 of maintaining the gauge invariance property.

\section{Acknowledgements}

This work is supported in part by INTAS grant \#99-1-590.

\end{document}